\documentstyle[twoside]{article}

\catcode`\@=11
\long\def\@makefntext#1{
\protect\noindent \hbox to 3.2pt {\hskip-.9pt  
$^{{\eightrm\@thefnmark}}$\hfil}#1\hfill}               

\def\@makefnmark{\hbox to 0pt{$^{\@thefnmark}$\hss}}    
        
\def\ps@myheadings{\let\@mkboth\@gobbletwo
\def\@oddhead{\hbox{}
\rightmark\hfil\eightrm\thepage}   
\def\@oddfoot{}\def\@evenhead{\eightrm\thepage\hfil
\leftmark\hbox{}}\def\@evenfoot{}
\def\sectionmark##1{}\def\subsectionmark##1{}}

\oddsidemargin=\evensidemargin
\newcounter{sectionc}\newcounter{subsectionc}\newcounter{subsubsectionc}
\renewcommand{\section}[1] {\vspace{12pt}\addtocounter{sectionc}{1} 
\setcounter{subsectionc}{0}\setcounter{subsubsectionc}{0}\noindent 
        {\tenbf\thesectionc. #1}\par\vspace{5pt}}
\renewcommand{\subsection}[1] {\vspace{12pt}\addtocounter{subsectionc}{1} 
        \setcounter{subsubsectionc}{0}\noindent 
        {\bf\thesectionc.\thesubsectionc. {\kern1pt \bfit #1}}\par\vspace{5pt}}
\renewcommand{\subsubsection}[1] {\vspace{12pt}\addtocounter{subsubsectionc}{1}
        \noindent{\tenrm\thesectionc.\thesubsectionc.\thesubsubsectionc.
        {\kern1pt \tenit #1}}\par\vspace{5pt}}
\newcommand{\nonumsection}[1] {\vspace{12pt}\noindent{\tenbf #1}
        \par\vspace{5pt}}

\newcounter{appendixc}
\newcounter{subappendixc}[appendixc]
\newcounter{subsubappendixc}[subappendixc]
\renewcommand{\thesubappendixc}{\Alph{appendixc}.\arabic{subappendixc}}
\renewcommand{\thesubsubappendixc}
        {\Alph{appendixc}.\arabic{subappendixc}.\arabic{subsubappendixc}}

\renewcommand{\appendix}[1] {\vspace{12pt}
        \refstepcounter{appendixc}
        \setcounter{figure}{0}
        \setcounter{table}{0}
        \setcounter{lemma}{0}
        \setcounter{theorem}{0}
        \setcounter{corollary}{0}
        \setcounter{definition}{0}
        \setcounter{equation}{0}
        \renewcommand{\thefigure}{\Alph{appendixc}.\arabic{figure}}
        \renewcommand{\thetable}{\Alph{appendixc}.\arabic{table}}
        \renewcommand{\theappendixc}{\Alph{appendixc}}
        \renewcommand{\thelemma}{\Alph{appendixc}.\arabic{lemma}}
        \renewcommand{\thetheorem}{\Alph{appendixc}.\arabic{theorem}}
        \renewcommand{\thedefinition}{\Alph{appendixc}.\arabic{definition}}
        \renewcommand{\thecorollary}{\Alph{appendixc}.\arabic{corollary}}
        \renewcommand{\theequation}{\Alph{appendixc}.\arabic{equation}}
        \noindent{\tenbf Appendix \theappendixc #1}\par\vspace{5pt}}
\newcommand{\subappendix}[1] {\vspace{12pt}
        \refstepcounter{subappendixc}
        \noindent{\bf Appendix \thesubappendixc. {\kern1pt \bfit #1}}
        \par\vspace{5pt}}
\newcommand{\subsubappendix}[1] {\vspace{12pt}
        \refstepcounter{subsubappendixc}
        \noindent{\rm Appendix \thesubsubappendixc. {\kern1pt \tenit #1}}
        \par\vspace{5pt}}

\newcommand{\textlineskip}{\baselineskip=13pt}
\newcommand{\smalllineskip}{\baselineskip=10pt}

\def\eightcirc{
\begin{picture}(0,0)
\put(4.4,1.8){\circle{6.5}}
\end{picture}}
\def\eightcopyright{\eightcirc\kern2.7pt\hbox{\eightrm c}} 

\newcommand{\copyrightheading}[1]
        {\vspace*{-2.5cm}\smalllineskip{\flushleft
        {\footnotesize International Journal of Modern Physics A, #1}\\
        {\footnotesize $\eightcopyright$\, World Scientific Publishing
         Company}\\
         }}

\def\abstracts#1#2#3{{
        \centering{\begin{minipage}{4.5in}\baselineskip=10pt\footnotesize
        \parindent=0pt #1\par 
        \parindent=15pt #2\par
        \parindent=15pt #3
        \end{minipage}}\par}} 

\newcommand{\bibit}{\nineit}

\renewenvironment{thebibliography}[1]
        {\frenchspacing
         \ninerm\baselineskip=11pt
         \begin{list}{\arabic{enumi}.}
        {\usecounter{enumi}\setlength{\parsep}{0pt}
         \setlength{\leftmargin 12.7pt}{\rightmargin 0pt} 
         \setlength{\itemsep}{0pt} \settowidth
        {\labelwidth}{#1.}\sloppy}}{\end{list}}

\newcounter{itemlistc}
\newcounter{romanlistc}
\newcounter{alphlistc}
\newcounter{arabiclistc}

\newcommand{\fcaption}[1]{
        \refstepcounter{figure}
        \setbox\@tempboxa = \hbox{\footnotesize Fig.~\thefigure. #1}
        \ifdim \wd\@tempboxa > 5in
           {\begin{center}
        \parbox{5in}{\footnotesize\smalllineskip Fig.~\thefigure. #1}
            \end{center}}
        \else
             {\begin{center}
             {\footnotesize Fig.~\thefigure. #1}
              \end{center}}
        \fi}

\newcommand{\tcaption}[1]{
        \refstepcounter{table}
        \setbox\@tempboxa = \hbox{\footnotesize Table~\thetable. #1}
        \ifdim \wd\@tempboxa > 5in
           {\begin{center}
        \parbox{5in}{\footnotesize\smalllineskip Table~\thetable. #1}
            \end{center}}
        \else
             {\begin{center}
             {\footnotesize Table~\thetable. #1}
              \end{center}}
        \fi}

\def\@citex[#1]#2{\if@filesw\immediate\write\@auxout
        {\string\citation{#2}}\fi
\def\@citea{}\@cite{\@for\@citeb:=#2\do
        {\@citea\def\@citea{,}\@ifundefined
        {b@\@citeb}{{\bf ?}\@warning
        {Citation `\@citeb' on page \thepage \space undefined}}
        {\csname b@\@citeb\endcsname}}}{#1}}

\newif\if@cghi
\def\cite{\@cghitrue\@ifnextchar [{\@tempswatrue
        \@citex}{\@tempswafalse\@citex[]}}
\def\citelow{\@cghifalse\@ifnextchar [{\@tempswatrue
        \@citex}{\@tempswafalse\@citex[]}}
\def\@cite#1#2{{$\null^{#1}$\if@tempswa\typeout
        {IJCGA warning: optional citation argument 
        ignored: `#2'} \fi}}

\def\pmb#1{\setbox0=\hbox{#1}
        \kern-.025em\copy0\kern-\wd0
        \kern.05em\copy0\kern-\wd0
        \kern-.025em\raise.0433em\box0}

\def\fnt#1#2{\footnotetext{\kern-.3em
        {$^{\mbox{\scriptsize #1}}$}{#2}}}

\def\fpage#1{\begingroup
\voffset=.3in
\thispagestyle{empty}\begin{table}[b]\centerline{\footnotesize #1}
        \end{table}\endgroup}

\font\tenrm=cmr10
\font\tenit=cmti10 
\font\tenbf=cmbx10
\font\bfit=cmbxti10 at 10pt
\font\ninerm=cmr9
\font\nineit=cmti9

\font\eightrm=cmr8

\def\qed{\hbox{${\vcenter{\vbox{                        
   \hrule height 0.4pt\hbox{\vrule width 0.4pt height 6pt
   \kern5pt\vrule width 0.4pt}\hrule height 0.4pt}}}$}}

\begin{document}


\normalsize\textlineskip
\thispagestyle{empty}
\setcounter{page}{1}

\copyrightheading{}                     
\vspace{-0.8cm}
\begin{flushright}
UNIL-IPT-00-24\\
hep-th/0010276\\
\end{flushright}

\vspace*{0.88truein}

\vspace{-0.9cm}

\fpage{1}
\centerline{\bf LOCALIZING GRAVITY ON A 3-BRANE}
\vspace*{0.035truein}
\centerline{\bf IN HIGHER DIMENSIONS}
\vspace*{0.37truein}
\centerline{{\footnotesize TONY GHERGHETTA}\footnote{Talk given at 
DPF 2000: The Meeting of the Division of Particles and Fields of the American 
Physical Society, Columbus, Ohio, 9-12 Aug 2000.}}
\vspace*{0.015truein}
\centerline{\footnotesize\it Institute of Theoretical Physics, University of Lausanne}
\baselineskip=10pt
\centerline{\footnotesize\it Lausanne, CH-1015,
Switzerland}

\vspace*{0.21truein}
\abstracts{We present metric solutions in six and higher dimensions with a bulk
cosmological constant, where gravity is localized on a 3-brane. The corrections 
to four-dimensional gravity from the bulk continuum modes are power-law suppressed. 
Furthermore, the introduction of a bulk ``hedgehog'' magnetic 
field leads to a regular geometry, and can localize gravity on the 3-brane with 
either positive, zero or negative bulk cosmological constant.}{}{}

\textlineskip                   
\vspace*{12pt}                  

\textheight=7.8truein

The idea that we live inside a domain wall in a five-dimensional (5d) 
universe,$^1$ has recently received new impetus. A solution of the 5d 
Einstein equations with a bulk cosmological constant has been found
where gravity is localized on a 3-brane.$^2$ The resulting 
four-dimensional (4d) Kaluza-Klein spectrum consists of a massless zero-mode 
localized at the origin, and a gapless continuous spectrum for the non-zero 
modes. Remarkably, one still recovers the usual 4d Newton's law at the origin.

This idea can be generalized to six and higher dimensions where gravity
is localized on a 3-brane which is a topological local defect of the 
higher-dimensional theory.$^{3,4}$ 
In D-dimensions the Einstein equations with a bulk cosmological
constant $\Lambda_D$  and stress-energy tensor $T_{AB}$ are
\begin{equation}
\label{eineqs}
    R_{AB} - \frac{1}{2} g_{AB} R = \frac{1}{M_D^{n+2}}\left(\Lambda_D
   g_{AB} + T_{AB}\right)~,
\end{equation}
where $M_D$ is the reduced D-dimensional Planck scale.  
A D-dimensional metric ansatz that repects 4d Poincare invariance
with $n$ transverse spherical coordinates satisfying $0\leq \rho < \infty$,  
$0\leq \{\theta_{n-1},\dots,\theta_2\} < \pi$ and $0\leq \theta_1
< 2\pi$, is
\begin{equation}
\label{metric}
    ds^2 = \sigma(\rho) g_{\mu\nu} dx^\mu dx^\nu 
    -d\rho^2-\gamma(\rho)d\Omega_{n-1}^2\, ,
\end{equation}
where  the metric signature of $g_{\mu\nu}$ is $(+,-,-,-)$ 
and $d\Omega_{n-1}^2$ is the surface area element.
At the origin $\rho=0$ we will assume that there is a core defect
of radius $\rho < \epsilon$, whose source is parameterised by the 
brane tension components, $\mu_0,\mu_\rho,$ and $\mu_\theta$.
In order to have a regular geometry at the origin we will require that the solution
satisfies
\begin{equation}
\label{bc}
     \sigma^\prime\big|_{\rho=0} = 0~, \quad (\sqrt{\gamma})^\prime
     \big|_{\rho=0} = 1 , \quad  \sigma\big|_{\rho=0} = A~, 
     \quad {\rm and} 
     \quad \gamma\big|_{\rho=0} = 0~,
\end{equation}
where $A$ is a constant. This leads to the following set of boundary conditions
\begin{equation}
\label{junc1}
     \sigma\sigma^\prime\sqrt{\gamma^{n-1}} \big|_0^\epsilon = 
       \frac{2}{(n+2)}\frac{1}{M_D^{n+2}}\bigg((n-2)\mu_0-\mu_\rho 
       -(n-1)\mu_\theta\bigg)~,
\end{equation}
and
\begin{equation}
\label{junc2}
     \sigma^2\sqrt{\gamma^{n-2}}(\sqrt{\gamma})^\prime \big|_0^\epsilon = 
       -\frac{1}{(n+2)}\frac{1}{M_D^{n+2}}\bigg(4\mu_0+\mu_\rho
     -3\mu_\theta\bigg)~,
\end{equation}
where it is understood that the limit $\epsilon\rightarrow 0$ is
taken. The equations (\ref{junc1}) and (\ref{junc2}) are the general
conditions relating the brane tension components to the metric
solution of the Einstein equations
(\ref{eineqs}),  and lead to nontrivial
relationships between the  components of the brane tension per unit
length.

For example, in six dimensions the solution outside the core defect
is given by$^3$
\begin{equation}
\label{expsoln}
      \sigma(\rho) = e^{-c \rho} \quad \quad {\rm and} \quad\quad 
         \gamma(\rho)= R_0^2  e^{-c \rho}~,
\end{equation}
where $c=\sqrt{\frac{2}{5}\frac{(-\Lambda_6)}{M_6^4}}$, and the arbitrary 
integration constant (which corresponds to an overall rescaling of 
the coordinates $x^\mu$), is chosen 
such that $\displaystyle \lim_{\epsilon\rightarrow 0} 
\sigma(\epsilon)=1$. Clearly the negative exponential solution requires that 
$\Lambda_6 < 0$. Thus, the geometry of the solution outside the core is simply 
$AdS_6/\Gamma$ where $\Gamma$ corresponds to a periodic identification of one
of the coordinates. 
If we now demand  that the solution (\ref{expsoln}) is consistent with the 
boundary conditions (\ref{junc1}) and (\ref{junc2}), 
the brane tension components must satisfy 
\begin{equation}
\label{cond1}
         \mu_0 = \mu_\theta + A^2 M_6^4~,
\end{equation}
where $\mu_\rho$ remains undetermined. Even though the condition (\ref{cond1})
amounts to a fine-tuning, one can imagine that it results from an underlying 
supersymmetry. This is precisely what happens in the supersymmetric
version of the 5d Randall-Sundrum solution.$^5$
Thus, it is encouraging to note that the bosonic background of the
6d solution (\ref{expsoln}) can also be supersymmetrized.$^6$

The Kaluza-Klein graviton spectrum of the 6d solution contains 
a zero-mode which is localized at the origin $\rho=0$. The nonzero modes 
consist of radial modes which form
a gapless continuum, together with the discrete angular modes with mass scale
$R_0^{-1}$. 
At distances $r\gg R_0$, the correction from the
bulk continuum radial modes to the 4d gravitational potential is
\begin{equation}
      V(r) = G_N  \frac{m_1 m_2}{r}  \left[1+ 
                  \frac{32}{3\pi}  \frac{1}{(c r)^3} \right]~.
\end{equation}
This compares with the result in 5d where the corrections from the 
continuum modes are ${\cal O}(1/r^2)$.

The spherically symmetric setup (\ref{metric}) can only be generalized 
to dimensions greater than six, provided that there are additional 
energy-momentum sources in the bulk.
A particularly interesting possibility is that due to $p$-form fields.$^4$
Consider the D-dimensional action
\begin{equation}
     S= \int d^Dx \sqrt{|g|} \left( \frac{1}{2} M_D^{n+2} R - 
     \frac{\Lambda_D}{M_D^{n+2}} + (-1)^p\frac{1}{4} 
      F_{\mu_1\dots\mu_{p+1}}F^{\mu_1\dots\mu_{p+1}} \right)~.
\end{equation}
When $p=n-2$, a solution to the equation of motion for the $p$-form field is
\begin{equation}
      F_{\theta_1\dots\theta_{n-1}} = Q
(\sin\theta_{n-1})^{(n-2)}\dots \sin\theta_2~,
\end{equation}
where $Q$ is the charge of the field configuration, and all other
components of $F$ are equal to zero. In fact, this ``hedgehog''
field configuration is the generalization of the magnetic field of a monopole,
and consequently the stability of these configurations is ensured by
magnetic flux conservation.
Outside the core $(\rho > \epsilon)$ we will assume a solution of the form
\begin{equation}
\label{pfansatz}
    \sigma(\rho)= e^{-c\rho} \quad {\rm and} \quad \gamma(\rho)
       = {\rm constant}~,
\end{equation}
where the arbitrary integration constant is again chosen such that
$\displaystyle \lim_{\epsilon\rightarrow 0} \sigma(\epsilon)=1$.
With this ansatz and including the contribution of the $p$-form bulk
field to the stress-energy tensor, the Einstein equations 
(\ref{eineqs}), are reduced to the following two equations for the metric 
factors outside the 3-brane source
\begin{eqnarray}
\label{pfsoln1}
    (n-1)!\frac{Q^2}{\gamma^{n-1}}-\frac{1}{2\gamma}(n-2)(n+2)
    +\frac{\Lambda_D}{M_D^{n+2}} =0~,    \\
\label{pfsoln2}
   c^2 = -\frac{1}{2} \frac{\Lambda_D}{M_D^{n+2}}
          + \frac{1}{4\gamma}(n-2)^2~.\qquad\qquad
\end{eqnarray}
We are interested in the solutions of these two algebraic equations which
lead to an exponential, $c^2>0$ and do not change the metric signature, 
$\gamma>0$. Remarkably, solutions to these equations 
exist for which these conditions can be simultaneously satisfied. 
In particular, for the $n=3$ case, there are solutions not only for 
$\Lambda_7< 0$, but also for $\Lambda_7 \geq0$, provided that 
$Q^2\Lambda_7/M_7^5 < 1/2$. Thus, the bulk cosmological constant does 
not need to be negative in order to localize gravity.$^4$ Similar 
solutions exist for all $n\geq 3$.

The equation of motion for the spin-2 radial modes using the solution
(\ref{pfansatz}) is qualitatively similar to the 5d case. The constant
$\gamma$ factor effectively plays no role in the localization of gravity.
Thus, the corrections to Newton's law will be suppressed by $1/r^2$
for all solutions $n\geq 3$. This is easy to understand since the
geometry outside the core defect is simply $AdS_5 \times S^{n-1}$, and 
there is just one non-compact dimension for all $n\geq 3$.

\nonumsection{Acknowledgements}
\noindent
This work was supported by the FNRS, contract no. 21-55560.98.

\nonumsection{References}

\end{document}